\documentclass[twocolumn,preprintnumbers,superscriptaddress,prl,nofootinbib]{revtex4-1}
\usepackage{graphicx}
\usepackage{times}
\usepackage{epsfig}
\usepackage{amsmath}
\usepackage{amsfonts}
\usepackage{amssymb}
\usepackage{url}
\usepackage{subfigure}
\usepackage{color}
\newcommand{\be}{\begin{equation}}
\newcommand{\ee}{\end{equation}}
\newcommand{\bea}{\begin{eqnarray}}
\newcommand{\eea}{\end{eqnarray}}
\def\tr{\mathrm{tr}}

\begin{document}
%\preprint{}
% Page numbers bottom-center
\pagestyle{plain}
%%%%%%%%%%%%%%%%%%%%%%%%%%%%%%%%%%%%%%%%%%%%%%%%%%%%%%%%%%%%
\title{
Scalar dark matter, Type II Seesaw and the DAMPE cosmic ray $e^+ + e^-$ excess
}
%%%%%%%%%%%%%%%%%%%%%%%%%%%%%%%%%%%%%%%%%%%%%%%%%%%%%%%%%%%%%
\author{Tong Li}
%\email{}
\affiliation{
School of Physics, Nankai University, Tianjin 300071, China
}
\affiliation{
ARC Centre of Excellence for Particle Physics at the Terascale, School of Physics and Astronomy,
Monash University, Melbourne, Victoria 3800, Australia
}
\author{Nobuchika Okada}
%\email{okadan@ua.edu}
\affiliation{
Department of Physics and Astronomy,
University of Alabama,
Tuscaloosa, AL 35487, USA
}
\author{Qaisar Shafi}
%\Email{shafi@bartol.udel.edu}
\affiliation{
Bartol Research Institute,
Department of Physics and Astronomy,
University of Delaware, Newark, DE 19716, USA
}
%\date{\today}

%\baselineskip 36pt

%%%%%%%%%%%%%%%%%%%%%%%%%%%%%%%%%%%%%%%%%%%%%%%%%%%%%%%%%%%%
\begin{abstract}

The DArk Matter Particle Explorer (DAMPE) has reported a measurement
  of the flux of high energy cosmic ray electrons plus positrons (CREs)
  in the energy range between $25$ GeV and $4.6$ TeV.
With unprecedented high energy resolution,
  the DAMPE data exhibit an excess of the CREs flux at an energy of around $1.4$ TeV.
In this letter, we discuss how the observed excess can be understood in a minimal framework
  where the Standard Model (SM) is supplemented by a stable SM singlet scalar as dark matter (DM)
  and type II seesaw for generating the neutrino mass matrix.
In our framework, a pair of DM particles annihilates into a pair of the SM SU(2) triplet scalars ($\Delta$s) in type II seesaw,
  and the subsequent $\Delta$ decays create the primary source of the excessive CREs around $1.4$ TeV.
The lepton flavor structure of the primary source of CREs has a direct relation with the neutrino oscillation data.
We find that the DM interpretation of the DAMPE excess determines
  the pattern of neutrino mass spectrum to be the inverted hierarchy type,
  taking into account the constraints from the Fermi-LAT observations of dwarf spheroidal galaxies.

\end{abstract}
\maketitle
%%%%%%%%%%%%%%%%%%%%%%%%%%%%%%%%%%%%%%%%%%%%%%%%
%\section{Introduction}
%%%%%%%%%%%%%%%%%%%%%%%%%%%%%%%%%%%%%%%%%%%%%%%%Sommerfeld

The DArk Matter Particle Explorer (DAMPE) has reported a new measurement
   of the flux of high energy cosmic ray electrons plus positrons (CREs)
   in the energy range of $25$ GeV-$4.6$ TeV \cite{DAMPE}.
This experiment has an excellent performance for the CREs energy resolution at the TeV scale
   and a high power of the hadron rejection.
The DAMPE data exhibit not only a spectral break around $0.9$ TeV,
   which was indicated by the H.E.S.S. experiments \cite{HESS},
   but also display an intriguing excess at an energy of around  $1.4$ TeV.

The DAMPE excess in the CREs spectrum has already stimulated a number of proposals
   for particle physics interpretations through pair annihilations
   of dark matter (DM) particles \cite{Fan:2017sor,DMpapers}.
Since the excess is localized around $1.4$ TeV, we may consider
   a pair of DM particles is mostly annihilating into leptons, namely, "leptophilic dark matter."
In Ref.~\cite{Yuan:2017ysv} (see also Ref.~\cite{Athron:2017drj}),
   the authors have performed a model-independent analysis
   to fit the DAMPE excess with a variety of leptonic channels ($\ell^+ \ell^-$ and $4 \ell$, where $\ell=e, \mu, \tau$)
   from DM annihilations or late-time DM decays,
   along with the constraints from the Fermi-LAT observations of dwarf spheroidal galaxies \cite{Fermi-LAT, Fermi-LAT2}
   and the Planck observations of Cosmic Microwave Background anisotropies \cite{Planck2015}.
It has been shown \cite{Yuan:2017ysv} that the Fermi-LAT observations disfavor the $\tau$ channels
   for the DM annihilation and the entire region for late-time DM decays.
In addition, the interpretation with DM annihilations invokes
   a `boost' factor which could either have  an astrophysical origin
   (large inhomogeneities in the dark matter distribution),
   or have some particle physics origin.

In this letter, we revisit a very simple extension of the SM
   in which two major missing pieces in the SM, namely, a dark matter candidate and the neutrino mass matrix, are incorporated.
The model was proposed some years ago \cite{GOS} to interpret an excess
   of cosmic ray positions reported by the PAMELA  experiment \cite{PAMELA}.
More detailed analysis for the cosmic ray fluxes was performed in Ref.~\cite{DGOS}.
The DM particle in our scheme is a SM singlet scalar $D$ \cite{Darkon},
   and its stability is ensured by an unbroken $Z_2$ symmetry
   under which it carries negative parity.
The leptophilic nature of this DM particle arises from its interactions
   with the SU(2) triplet scalar field ($\Delta$) which is introduced
   to accommodate the observed neutrino oscillations \cite{PDG}
   via the type II seesaw mechanism \cite{typeII}.

%%%%%%%%%%%%%%%%%%%%%%%%%%%%%%%%%%%%%%%%%%%%%%%
\begin{table}[t]
\begin{center}
\begin{tabular}{c|cc|c}
           & SU(2)$_L$ & U(1)$_Y$ & $Z_2$  \\
\hline
$ \ell_L^i   $ & {\bf 2}  & $-1/2$    & $+$  \\
$ H      $ & {\bf 2}  & $+1/2$    & $+$  \\
\hline
$ \Delta $ & {\bf 3}  & $+1$      & $+$  \\
\hline
$ D      $ & {\bf 1}  & $ 0  $    & $-$  \\
\end{tabular}
\end{center}
\caption{
Particle content relevant for our discussion in this letter.
In addition to the SM lepton doublets $\ell_L^i$ ($i=1,2,3$ being the generation index)
 and the Higgs doublet $H$,
 a complex scalar $\Delta$ and a real scalar $D$ are introduced.
The SM SU(2)$_L$ triplet scaler $\Delta$ plays the key role in the type II seesaw mechanism,
  while $D$ is the DM candidate.
}
\label{Tab:1}
\end{table}
%%%%%%%%%%%%%%%%%%%%%%%%%%%%%%%%%%%%%%%%%%%%%%%

The particle content relevant for our discussion in this letter is summarized  in Table~\ref{Tab:1}.
An odd  $Z_2$ parity is assigned to the SM singlet scalar ($D$), which makes it stable and a suitable DM candidate.
It is often useful to explicitly express the triplet scalar by three complex scalars
 (electric charge neutral ($\Delta^0$), singly charged ($\Delta^+$) and doubly charged ($\Delta^{++}$) scalars):
\bea
 \Delta=\frac{\sigma^i}{\sqrt{2}}
   \Delta_i=\left(
 \begin{array}{cc}
    \Delta^+/\sqrt{2} & \Delta^{++}\\
    \Delta^0 & -\Delta^+/\sqrt{2}\\
 \end{array}\right) ,
\eea
where $\sigma^i$'s are Pauli matrices.

Following the notations of Ref.~\cite{Schmidt:2007nq}, the scalar potential relevant for type II seesaw is given by
\bea
 V(H, \Delta) &=&
 -m_H^2 (H^\dagger H)
  + \frac{\lambda}{2} (H^\dagger H)^2
  \nonumber\\
&+& M_\Delta^2 \, \tr \left[ \Delta^\dagger \Delta \right]
 + \frac{\lambda_1}{2} \left( \tr [\Delta^\dagger \Delta] \right)^2
 \nonumber\\
&+& \frac{\lambda_2}{2}\left(
 \left( \tr [\Delta^\dagger \Delta] \right)^2
 - \tr \left[  \Delta^\dagger \Delta \Delta^\dagger \Delta \right]
  \right)
 \nonumber\\
&+&
 \lambda_4 H^\dagger H \; \tr \left(\Delta^\dagger\Delta\right)
 + \lambda_5 H^\dagger
 \left[\Delta^\dagger, \Delta\right] H
\nonumber \\
&+& \left[ 2 \lambda_6 M_\Delta
    H^T i\sigma_2 \Delta^\dagger H +{\rm H.c.} \right],
 \label{H-Delta-Potential}
\eea
where the coupling constants $\lambda_i$
 are taken to be real without loss of generality.
The triplet scalar ($\Delta$) has a Yukawa coupling with the lepton doublets given by
\bea
{\cal L}_\Delta &=&
 -\frac{1}{\sqrt{2}}\left(Y_\Delta\right)_{ij}
  \ell_L^{Ti}\, \mathrm{C} \, i\, \sigma_2\,  \Delta \, \ell_L^j +{\rm H.c.}\nonumber \\
  &=&
 -\frac{1}{\sqrt{2}} \left(Y_\Delta\right)_{ij} \,
  \nu^{Ti}_L \, \mathrm{C} \, \Delta^0 \, \nu_L^j
 \nonumber\\
  &+&  \frac{1}{2} \, \left(Y_\Delta\right)_{ij} \,
  \nu^{Ti}_L \, \mathrm{C} \, \Delta^+ \ e_L^j
 \nonumber\\
  &+& \frac{1}{\sqrt{2}} \left(Y_\Delta\right)_{ij} \, e^{Ti}_L \,
  \mathrm{C} \, \Delta^{++} \  e_L^j  +  \text{H.c.}  ,
\label{Yukawa}
\eea
where $\mathrm{C}$ is the charge conjugate matrix,
 and $\left(Y_\Delta\right)_{ij}$ denotes the elements
 of the Yukawa matrix.

A non-zero vacuum expectation value (VEV) of the Higgs doublet
  generates a tadpole term for $\Delta$ through the last term in Eq.~(\ref{H-Delta-Potential}).
A non-zero VEV of the triplet Higgs is generated,
  $\langle \Delta^0 \rangle = v_\Delta/\sqrt{2} \simeq \lambda_6 v^2/M_\Delta$,
  from minimizing the scalar potential.
As a result, lepton number is spontaneously broken by $\Delta$.
From Eq.~(\ref{Yukawa}), this leads to the neutrino mass matrix:
\bea
  m_\nu = v_\Delta \, \left(Y_\Delta\right)_{ij} .
\eea
Here $v$ is the SM Higgs doublet VEV with $v^2+v_\Delta^2=(246 \ {\rm GeV})^2$.

Note that the triplet Higgs VEV contributes to the weak boson masses and
  alters the $\rho$-parameter from the SM prediction, $\rho = 1$, at tree level.
The current precision measurement \cite{PDG} constrains this deviation to be within the range,
 $ \Delta \rho =\rho -1 \simeq v_\Delta /v  \lesssim 0.01$,
 so that we obtain $\lambda_6 \lesssim 0.01 M_\Delta/v$.

We can fix the structure of $\left(Y_\Delta\right)_{ij}$ by using the neutrino oscillation data:
\bea
   Y_\Delta = \frac{1}{v_\Delta} U_{\rm MNS}^* D_\nu U_{\rm MNS}^\dagger,
\eea
where $U_{\rm MNS}$ is the neutrino mixing matrix in the standard form \cite{PDG},
  and $D_\nu={\rm diag}(m_1, m_2, m_3)$ is the neutrino mass eigenvalue matrix.
We employ the neutrino oscillation data:
   $\sin^2 2 \theta_{13}=0.092$ \cite{An:2012eh}
   along with $\sin^2 2 \theta_{12} = 0.87$, $\sin^2 2 \theta_{23} = 1.0$,
   $\Delta m_{12}^2 =m_2^2 -m_1^2=7.6 \times 10^{-5}$ eV$^2$,
   and $|\Delta m_{23}^2| =|m_3^2 -m_2^2| = 2.4 \times 10^{-3}$ eV$^2$ \cite{PDG}.
Motivated by the recent measurement of the Dirac $CP$-phase ($\delta_{CP}$),
  we set $\delta_{CP}=3 \pi/2$ \cite{T2K}.
For simplicity, we choose the lightest neutrino mass to be zero.
With $Y_\Delta \gtrsim v_\Delta/M_\Delta$, the triplet scalar $\Delta$
  dominantly decays to a pair of leptons, such as
  $\nu^i \nu^j$, $\nu^i e^j$, $e^i \nu^j$ and $e^i e^j$.
With the oscillation data inputs, we calculate the ratio for the charged lepton flavors produced
  by the $\Delta$ decay ($\Delta \to \nu^i e^j$, $e^i \nu^j$, $e^i e^j$) to be
\bea
 &&  e: \mu : \tau \simeq 0.1 : 1 : 1   \; \; ({\rm Normal \; \; Hierarchy}),
 \label{NH}
 \\
 &&  e: \mu : \tau \simeq 2 : 1 : 1   \; \; ({\rm Inverted \; \; Hierarchy}),
\label{IH}
\eea
%\Tong{two benchmarks in CMS PAS HIG-16-036:
%\bea
% &&  ee: \mu\mu : \tau\tau \simeq 0.01 : 1 : 1   \; \; ({\rm Normal \; \; Hierarchy}),  \nonumber \\
% &&  ee: \mu\mu : \tau\tau \simeq 4 : 1 : 1   \; \; ({\rm Inverted \; \; Hierarchy}),
%\label{nmass}
%\eea
%Do we need to add a comment on the flavor-off-diagonal dilepton processes $e\mu$, $e\tau$, $\mu\tau$ here?
%}
for the neutrino mass patterns of normal hierarchy and inverted hierarchy, respectively.
We find that these ratios are independent of the Majorana phases in $U_{\rm MNS}$.

The scalar potential relevant for DM physics is given by
\bea
&& V(H, \Delta, D)   \nonumber \\
&=& \frac{1}{2} m_0^2 D^2 + \lambda_D D^4
 + \lambda_H D^2  (H^\dagger H)
 + \lambda_\Delta D^2 \tr (\Delta^\dagger \Delta)  \nonumber \\
&=& \frac{1}{2} m_D^2 D^2 + \lambda_D D^4
 + \lambda_H v D^2 h  + \frac{\lambda_H}{2} D^2 h^2 \nonumber \\
&+& \lambda_\Delta D^2
\left(
 \sqrt{2} v_\Delta {\rm Re}[\Delta^0] + |\Delta^0|^2 + |\Delta^+|^2 + |\Delta^{++}|^2
 \right),
\label{H-Delta-D-Potential}
\eea
where $m_D^2 = m_0^2 + \lambda_H v^2 + \lambda_\Delta v_\Delta^2$
  is the DM mass, and $h$ is the physical Higgs boson.
Through the couplings $\lambda_H$ and $\lambda_\Delta$ in this scalar potential,
  a pair of DM particles annihilates into pairs of the Higgs doublet and the triplet,
  $DD \to H^\dagger H, \; \Delta^\dagger \Delta$.\footnote{
Although there are other DM annihilation processes such as $DD \to h \to W^+W^-$,
  they are subdominant, since we choose $M_\Delta \gg v$ in the following.
}
%\Tong{Nobuchika explains why the quartic coupling diagram is dominant over the Higgs portal process in s-channel and
%the neutral Higgs pair production via D in t-channel. Can we add a comment here?}
%
In order to evaluate the thermal DM relic abundance,
   we first calculate the thermally averaged cross section times relative velocity
   for the process in the non-relativistic limit, which is given by
 \bea
  \langle  \sigma v_{\rm rel} \rangle  &=&
   \frac{1}{16 \pi m_D^2}
 \left( \lambda_H^2  \sqrt{1- \frac{m_h^2}{m_D^2}}
      + 6 \lambda_\Delta^2  \sqrt{1- \frac{M_\Delta^2}{m_D^2}} \right) \nonumber \\
&\simeq &
   \frac{1}{16 \pi m_D^2}
 \left( \lambda_H^2  + 0.27  \lambda_\Delta^2   \right) ,
\label{sigmav}
\eea
where we have chosen $m_h=125$ GeV for the Higgs boson mass, $m_D - M_\Delta=3$ GeV with $m_D=3$ TeV.
In particular, the cross section of $DD$ annihilation into doubly and singly charged Higgs is
\begin{eqnarray}
&&  \langle  \sigma v_{\rm rel} \rangle(DD\to \Delta^{++}\Delta^{--}, \Delta^+\Delta^-)=\frac{1}{8 \pi m_D^2}\lambda_\Delta^2  \sqrt{1- \frac{M_\Delta^2}{m_D^2}}.\nonumber \\
\end{eqnarray}

The present DM relic density is determined by solving the Boltzmann equation
  with the thermally averaged cross section in Eq.~(\ref{sigmav}).
We employ an approximation formula given by \cite{Kolb:1990vq}
\begin{eqnarray}
 \Omega h^2
 &=&
 \frac{1.07\times 10^9 \, x_f{\rm GeV}^{-1}}
 {\sqrt{g_*}M_{\rm Pl}\langle\sigma v_{\rm rel}\rangle} ,
\label{Omega}
\end{eqnarray}
  where $M_{\rm Pl}=1.22 \times 10^{19}$ GeV is the Planck mass,
  the freeze-out temperature $x_f = m_D/T_f$ is given
  by  $x_f = \ln[X] - 0.5\ln[\ln[X]]$
  with $X = 0.038 (1/g_*^{1/2})M_{\rm Pl} m_D \langle \sigma v_{\rm rel}\rangle$, and
  we take $g_* =100$ for the total degrees of freedom of the thermal plasma.
We numerically find the solution,
\bea
    \lambda_H^2 + 0.27 \lambda_\Delta^2 \simeq 0.82,
\label{lamda}
\eea
  to reproduce the observed DM relic density \cite{Planck2015}
\bea
  \Omega_{\rm DM} h^2 \simeq 0.12.
\eea
The couplings in perturbative regime can satisfy Eq.~(\ref{lamda}).

\begin{figure}[h!]
\begin{center}
\includegraphics[scale=1,width=8cm]{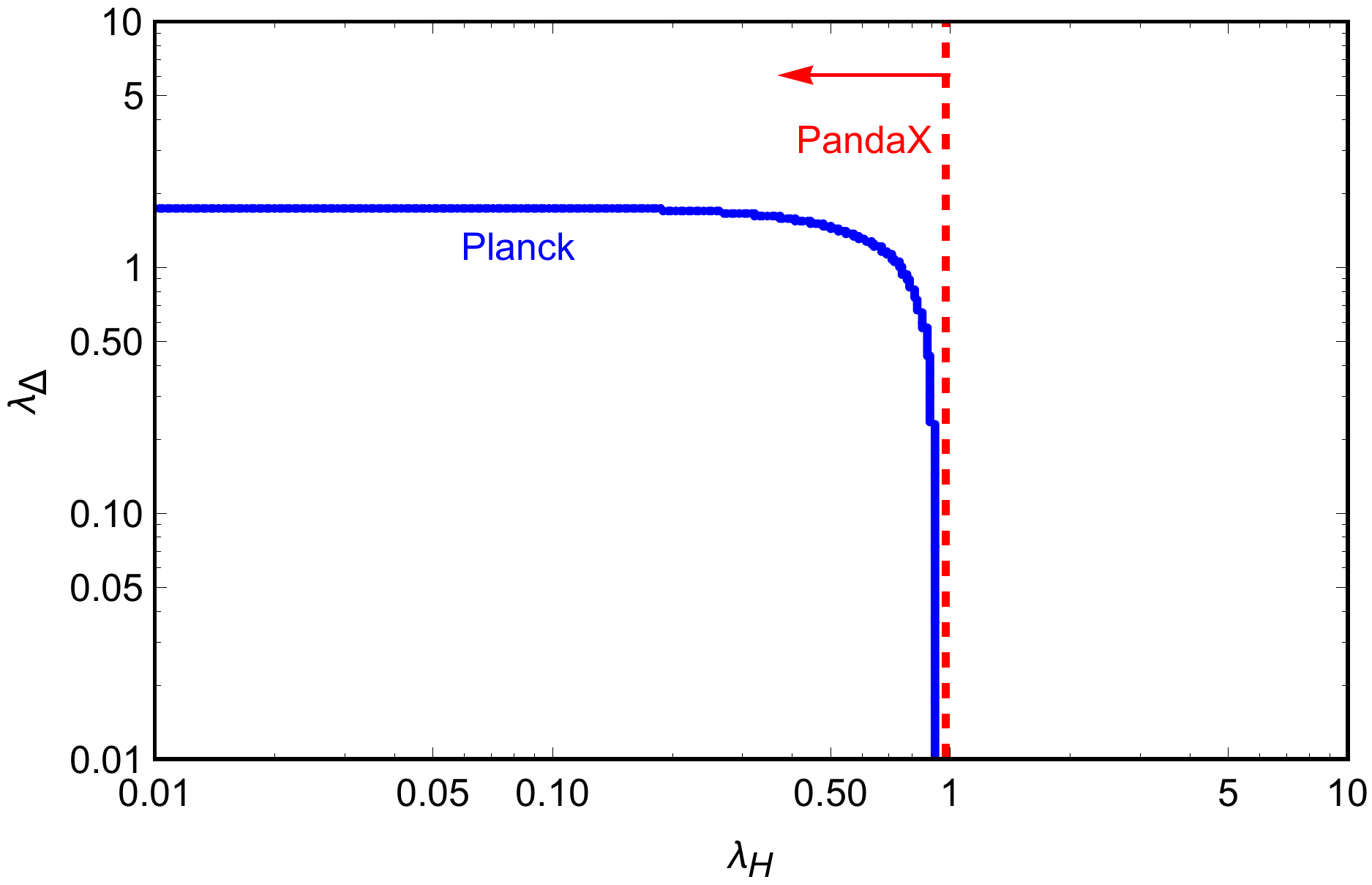}
\end{center}
\caption{Allowed $\lambda_\Delta$ vs. $\lambda_H$ by Planck and PandaX. We assume $m_D=3$ TeV and $m_D-M_\Delta=3$ GeV.}
\label{lambda}
\end{figure}

Several experiments are underway to directly detect the DM particles through elastic scattering off nuclei.
The most stringent limit on the spin-independent elastic scattering cross section has been obtained
  by the recent PandaX-II \cite{PandaX} experiment:
  $\sigma_{el} \leq 3 \times 10^{-9}$ pb for a DM mass of $m_D =3$ TeV.
This result leads to an upper bound on $\lambda_H$,
  since the scalar DM particle can scatter off a nucleon
  through processes mediated by the SM Higgs boson in the $t$-channel.
The spin-independent elastic scattering cross section is given by
\bea
\sigma_{{\rm SI}} =
 \frac{\lambda_H^2}{\pi m_h^4}
  \frac{m_N^2}{(m_N+m_D)^2}
  f_N^2,
\label{DD}
\eea
where $m_N=0.939$ GeV is the nucleon mass, and
\bea
f_N =
\left(
\sum_{q=u,d,s} f_{T_q} + \frac{2}{9}f_{TG}
\right)
m_N ,
\eea
   is the nuclear matrix element accounting for the quark and gluon contents of the nucleon.
In evaluating $f_{T_q}$, we employ the results from the lattice QCD simulation \cite{LatQCD}:
   $f_{T_u} +f_{T_d} \simeq 0.056$ and $|f_{T_s}|\leq 0.08$.
To make our analysis conservative, we set $f_{T_s}=0$.
Using the trace anomaly formula, $\sum_{q=u,d,s} f_{T_q} + f_{TG}=1$ \cite{TraceAnomaly},
   we obtain $f_N^2 = 0.0706 \, m_N^2$, and hence the spin-independent elastic scattering cross section
   is approximately given by
\bea
 \sigma_{{\rm SI}} =
3.2 \times 10^{-9} \; {\rm pb} \times \lambda_H^2 ,
\label{DD2}
\eea
  for $m_h=125$ GeV and $m_D=3$ TeV.
Hence, we find $\lambda_H^2 \leq 0.95$, which is not a strong constraint.
In Fig.~\ref{lambda} we show the allowed region of $\lambda_\Delta$ and $\lambda_H$ by Planck and PandaX for $m_D=3$ TeV and $m_D-M_\Delta=3$ GeV.

The dark matter in the halo of  our galaxy can annihilate into the Higgs doublet and triplet.
We choose $\lambda_H \ll \lambda_\Delta$, so that a pair of DM particles mainly annihilates into
   a pair of $\Delta$, followed by the decay $\Delta \to \ell^i \ell^j$.
In this way, the DM pair annihilations can produce 2 or 4 charged leptons.
We recall here that due to the constraints from the Fermi-LAT observations,
   the $\tau$ channel is disfavored as a dominant final state.
Therefore, the neutrino mass spectrum must exhibit inverted hierarchy from Eqs.~(\ref{NH}) and (\ref{IH}).
This is one of the main conclusions of this letter.
To explain the DAMPE excess, we expect $m_D \simeq M_\Delta$,
  so that each final state lepton has almost a line spectrum.
According to the fit results along with the Fermi-LAT constraints in Ref.~\cite{Yuan:2017ysv},
  we set $m_D \simeq 3$ TeV and a small mass differences between $D$ and $\Delta$,
  as we have chosen in Eq.~(\ref{sigmav}),
  to yield the energy of each lepton to be $E_\ell \simeq 1.5$ TeV.
Our choice of $M_\Delta \simeq 3$ TeV satisfies the current lower bound \cite{Aaboud:2017qph}, $M_\Delta \gtrsim 770$-$870$ GeV,
  from the search for doubly charged Higgs bosons at the Large Hadron Collider.
To account for the DAMPE excess, the DM annihilation cross section is found
  to be much larger than the thermal cross section \cite{Yuan:2017ysv},
  and some enhancement mechanism is necessary.
In the minimal version of our scenario, we simply assume that
  such an enhancement (boost factor) originates from large inhomogeneities
  in the dark matter distribution.

The cosmic ray propagation is described by the following transport equation~\cite{Ginzburg}
\begin{eqnarray}
\partial_tf-\partial_E(b(E)f)-D(E)\nabla^2f=Q,
\label{diffusion}
\end{eqnarray}
where $f$ is the density of cosmic rays, $b(E)=b_0(E/{\rm GeV})^2$ is the energy loss coefficient, $D(E)=D_0(E/{\rm GeV})^\delta$, and $Q=Q(\vec{x},E,t)$ is the source term.
We take $b_0=10^{-16} \ {\rm GeV/s}$, $D_0=11 \ {\rm pc^2/kpr}$, and $\delta=0.7$. In the steady-state case with only space diffusion and energy loss, the above equation can be solved in terms of the Green function~\cite{Kuhlen:2009is,Delahaye:2010ji}
\begin{eqnarray}
G(\vec{x},E;\vec{x}_s,E_s)={{\rm exp}[-(\vec{x}-\vec{x}_s)^2/\lambda^2]\over b(E)(\pi \lambda^2)^{3/2}},
\end{eqnarray}
with the propagation scale $\lambda$ given by
\begin{eqnarray}
\lambda^2=4\int_E^{E_s}dE'{D(E')\over b(E')}.
\end{eqnarray}
The solution of Eq.~(\ref{diffusion}) is then given by
\begin{eqnarray}
f(\vec{x},E)=\int d^3x_s \int dE_s G(\vec{x},E;\vec{x}_s,E_s)Q(\vec{x}_s,E_s).
\label{solution}
\end{eqnarray}
Finally, the electron/positron flux is $\Phi(\vec{x},E)=vf(\vec{x},E)/(4\pi)$ with $v$ being the cosmic ray velocity.

In Eq.~(\ref{solution}), the dark matter source term of electrons/positrons can be described by the product of the spatial distribution and the spectrum function
\begin{eqnarray}
Q(\vec{x},E)={1\over 2}{\rho^2(\vec{x})\over m_D^2}\langle \sigma v_{\rm rel} \rangle {dN\over dE},
\end{eqnarray}
where $\rho(\vec{x})$ is the DM spatial distribution, $\langle \sigma v_{\rm rel} \rangle$ is the total velocity averaged dark matter annihilation cross section, and $dN/ dE$ is the energy spectrum of cosmic ray particle produced in the annihilation. For the DM spatial distribution, we assume a generalized Navarro-Frenk-White (NFW) profile~\cite{NFW} to describe a DM subhalo with $d_s=0.3$ kpc distance away from us
\begin{eqnarray}
\rho(r)=\rho_s{(r/r_s)^{-\gamma}\over (1+r/r_s)^{3-\gamma}},
\end{eqnarray}
with $\gamma=0.5$ and $r_s=0.1$ kpc.

For the 4-body spectrum of $e^++e^-$ we consider, one has
\begin{eqnarray}
%{dN\over dE} &=& {\langle \sigma v\rangle_{\Delta^{\pm\pm}}\over \langle \sigma v\rangle} {\rm BR}^2(\Delta^{\pm\pm}\to e^\pm e^\pm)\frac{d\bar{N}}{dE},
%&&{dN\over dE} = {\langle \sigma v\rangle_{\Delta^{\pm\pm}}\over \langle \sigma v\rangle} 2\sum_{i,j}{\rm BR}(\Delta^{\pm\pm}\to i^\pm j^\pm)\left(\frac{d\bar{N}_i}{dE}+\frac{d\bar{N}_j}{dE}\right),\nonumber \\
%&&\approx {\langle \sigma v\rangle_{\Delta^{\pm\pm}}\over \langle \sigma v\rangle} 4\times {\rm BR}(\Delta^{\pm\pm}\to e^\pm e^\pm)\frac{d\bar{N}_e}{dE}, \ \ \ i,j=e,\mu,\tau
&&{dN\over dE} \approx {\langle \sigma v_{\rm rel}\rangle_{\Delta^{\pm\pm}}\over \langle \sigma v_{\rm rel} \rangle} 4\times {\rm BR}(\Delta^{\pm\pm}\to e^\pm e^\pm)\frac{d\bar{N}}{dE},
\label{4body}
\end{eqnarray}
where $\langle \sigma v_{\rm rel} \rangle_{\Delta^{\pm\pm}}/\langle \sigma v_{\rm rel}\rangle=\langle \sigma v_{\rm rel}\rangle(DD\to \Delta^{++}\Delta^{--})/\langle \sigma v_{\rm rel}\rangle \approx 1/3$ if $\lambda_H\ll \lambda_\Delta$ in our model and ${\rm BR}(\Delta^{\pm\pm}\to e^\pm e^\pm)=50\% (1\%)$ is the branching ratio of doubly charged triplet Higgs decay to same sign electrons/positrons in the case of inverted hierarchy (normal hierarchy). Note that in Eq.~(\ref{4body}) we ignore the decays with $\mu^\pm$ and $\tau^\pm$ in final states which give soft secondary electrons/positrons and are thus disfavored by DAMPE data~\cite{Yuan:2017ysv}. The cosmic ray spectrum $d\bar{N}/dE$ in the lab frame is given by the spectrum from the triplet Higgs decay in its rest frame, denoted by $dN/dE_0$, after a Lorentz boost~\cite{Elor:2015tva,Elor:2015bho}. Namely,
\begin{eqnarray}
{d\bar{N}\over dE}&=&\int^{t_{1,\rm max}}_{t_{1,\rm min}}{dx_0\over x_0\sqrt{1-\epsilon^2}}{dN\over dE_0},
\end{eqnarray}
where
\begin{eqnarray}
t_{1,\rm max}&=&{\rm min}\left[1,{2x\over \epsilon^2}\left(1+\sqrt{1-\epsilon^2}\right)\right],\\
t_{1,\rm min}&=&{2x\over \epsilon^2}\left(1-\sqrt{1-\epsilon^2}\right)
\end{eqnarray}
with $\epsilon=M_{\Delta}/m_D$ and $x=E/m_D\leq 0.5$.
We use PPPC4DM ID~\cite{Cirelli:2010xx} to generate the above energy spectrum of $e^++e^-$.

For the cosmic ray background not from DM contribution, we adopt a power law parameterization with two breaks,
\begin{eqnarray}
&&\Phi_{\rm bkg}=\nonumber \\
&&\Phi_0 E^{-\Delta\gamma}\left[1+\left({E_{\rm br,1}\over E}\right)^{\delta_0}\right]^{\Delta\gamma_1/\delta_0}\left[1+\left({E\over E_{\rm br,2}}\right)\right]^{\Delta\gamma_2/\delta_0}.\nonumber \\
\end{eqnarray}
Given the fit to the DAMPE data without the excess point and two fixed parameters, i.e. $E_{\rm br,1}=50$ GeV and $\delta_0=10$, one can obtain the other parameters as $\Phi_0=247.2 \ {\rm GeV}^{-1} \ {\rm m}^{-2} \ {\rm s}^{-1} \ {\rm sr}^{-1}$, $\Delta\gamma=3.092$, $\Delta\gamma_1=0.096$, $\Delta\gamma_2=-0.968$, and $E_{\rm br,2}=885.4$ GeV~\cite{Fan:2017sor}.

\begin{figure}[h!]
\begin{center}
\includegraphics[scale=1,width=8cm]{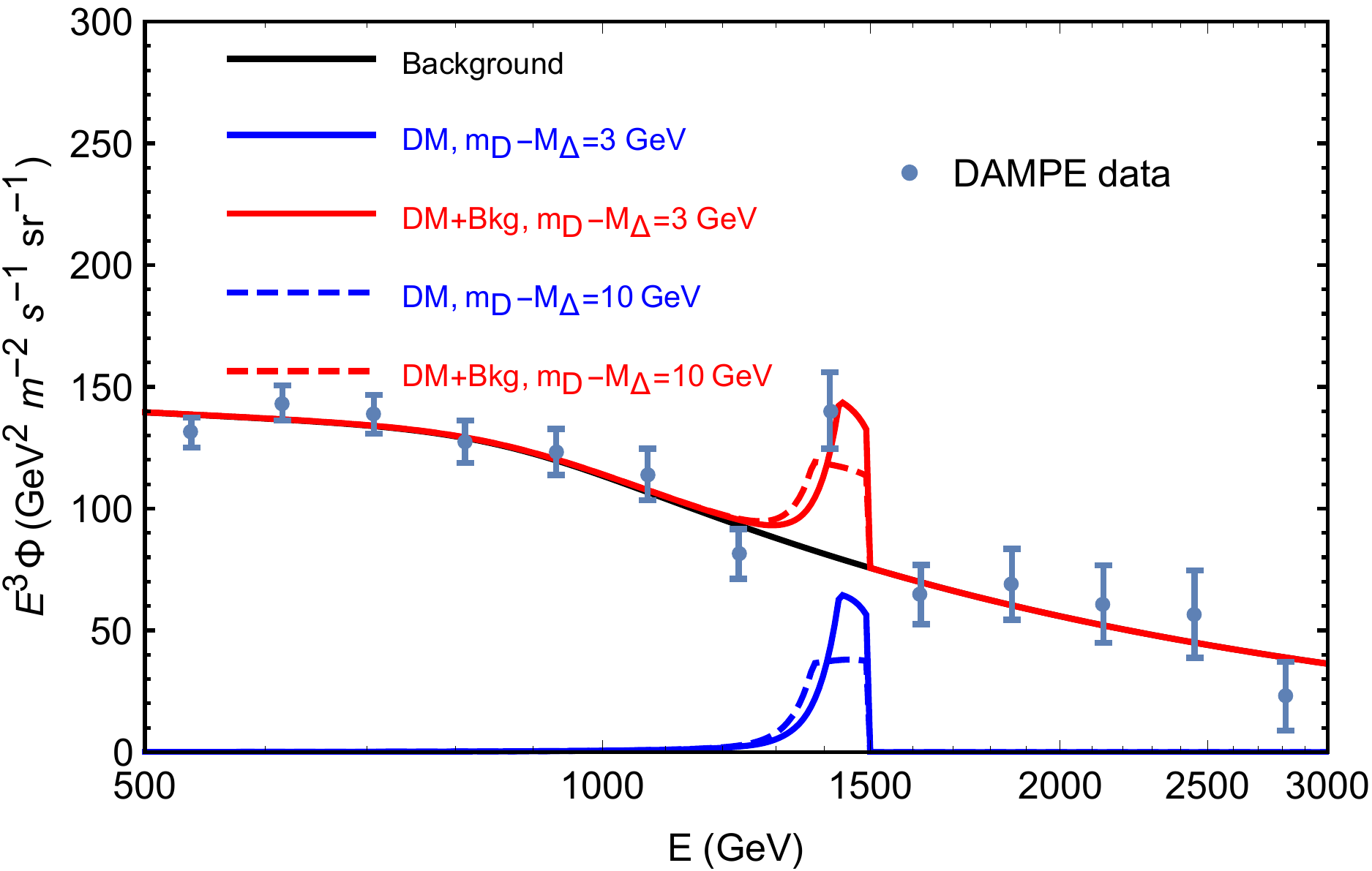}
\end{center}
\caption{The DAMPE $e^++e^-$ data and the DM signal from $DD\to \Delta^{++}\Delta^{--}\to e^+e^+e^-e^-$ with $m_D=3$ TeV and mass difference $m_D-M_\Delta=3$ GeV and 10 GeV. The fitted background is presented in black curve.}
\label{fitIH}
\end{figure}

Taking $\sigma v_{\rm rel}=3\times 10^{-26} \ {\rm cm}^3/{\rm s}$, $m_D=3$ TeV and the local density of $\rho_s=175 \ {\rm GeV}/{\rm cm}^3$, in Fig.~\ref{fitIH} we show the DAMPE data and the DM contributions to the $e^++e^-$ flux for the inverted hierarchy of neutrino mass pattern. The mass difference $m_D-M_\Delta$ is respectively assumed to be 3 GeV and 10 GeV. As expected, a smaller mass difference leads to a sharper energy spectrum and is more likely to explain the DAMPE excess. In the case of normal hierarchy, one needs an enhancement factor of about $0.5/0.01=50$ to fit the DAMPE data. This indicates that the inverted hierarchy spectrum is more preferred by the DAMPE excess.

When in addition including the DM annihilation into singly charged Higgs pairs with two electron/positron in final states, as $\sigma v_{\rm rel}(DD\to \Delta^{+}\Delta^{-})=\sigma v_{\rm rel}(DD\to \Delta^{++}\Delta^{--})$ if triplet Higgses being degenerate and ${\rm BR}(\Delta^{\pm}\to e^\pm \nu)\approx {\rm BR}(\Delta^{\pm\pm}\to e^\pm e^\pm)$ in both normal hierarchy and inverted hierarchy, we find the local density of $\rho_s\approx 140 \ {\rm GeV}/{\rm cm}^3$ is needed to get agreement with the DAMPE data.

In the above analysis, we assume the triplet Higgses are degenerate with the mass being $M_\Delta$. Actually, if the coupling $\lambda_5$ is sizable enough, the mass differences between triplet Higgses are likely to be larger than the sub-GeV mass splitting $m_D-M_{\Delta}$ required by DAMPE excess. Thus, the doubly charged Higgs can serve as the lightest triplet scalar with sub-GeV smaller mass than the DM, while the singly and neutral triplet Higgses are all heavier than DM. The DM particles subsequently annihilate into doubly charged Higgs pairs only, i.e. $\langle \sigma v_{\rm rel}\rangle_{\Delta^{\pm\pm}}/\langle \sigma v_{\rm rel}\rangle=1$ in Eq.~(\ref{4body}). In this case the local density is required to be $\rho_s\approx 100 \ {\rm GeV}/{\rm cm}^3$ to fit the DAMPE excess for $m_D-M_{\Delta^{\pm\pm}}=3$ GeV.

We may consider a particle physics origin for boosting the annihilation cross section
  of the DM particles.
In Ref.~\cite{GOS}, a real scalar $S$, which is $Z_2$-parity even
  and singlet under the SM gauge group, is introduced,
  and its mass is tuned to realize the Breit-Wigner enhancement
  for DM annihilations proposed in Ref.~\cite{Ibe:2008ye}. 
Specifically, we introduce the following scalar potential: 
\bea
 V(S, D, \Delta) = \frac{1}{2} M_S^2 S^2
           + \lambda_1 M_S S D^2
           + \lambda_2 M_S S \tr(\Delta^\dagger \Delta). \nonumber \\
\eea
  where we have parametrized the triple scalar couplings by the scalar mass $M_S$. 
For simplicity, we assume that other couplings involving $S$ are negligibly small. 
In the zero-velocity limit, the annihilation cross section of a process mediated
   by the singlet, $DD \to S \to \Delta^\dagger \Delta$,  
   is calculated to be
\bea
  \sigma v_{\rm rel}|_{v_{\rm rel} \to 0} = \frac{8 \lambda_1^2 M_S^2}{(4 m_D^2 - M_S^2)^2 + M_S^2
  \Gamma_S^2} \frac{\tilde{\Gamma}_S}{2 m_D},
\eea
where the total decay width of the $S$ boson is given by
 $\Gamma_S = (3 \lambda_2^2/16 \pi) M_S$,
 and $\tilde{\Gamma}_S= \Gamma_S(M_S \to 2 m_D)$.
According to Ref.~\cite{Ibe:2008ye}, we introduce
 two small parameters ($0 < \delta \ll 1$ and $\gamma \ll 1$) 
 defined with 
\bea
  M_S^2 = 4 m_D^2 (1-\delta), \; \;
  \gamma = \frac{\Gamma_S}{M_S}= \frac{3 \lambda_2^2}{16 \pi}. 
\eea
The cross section is then rewritten as
\bea
  \sigma v_{\rm rel}|_{v_{\rm rel} \to 0} \simeq \frac{2 \lambda_1^2}{m_D^2}
    \frac{\gamma}{\delta^2 + \gamma^2}.
\eea
For $\delta, \gamma \ll 1$, we have an enhancement of the annihilation cross section 
at the present universe \cite{Ibe:2008ye}. 
Although the same process is also relevant for DM annihilation in the early universe,
  a relative velocity $v_{\rm rel} = {\cal O}(0.1)$ is not negligible, and the total energy 
  of annihilating DM particles is away from the resonance pole.
As a result, the annihilation cross section at the freeze-out time is suppressed,
  compared to the one at present.

In this mechanism, the DM pair annihilation process in the present universe
  is enhanced through the $s$-channel resonance with an intermediate state.
Although the same process is also relevant for DM annihilation in the early universe,
  a relative velocity between annihilating DM particles at the freeze-out time
  is not negligible, and the total energy of annihilating DM particles is away from
  the $s$-channel resonance pole.
As a result, the annihilation cross section at the freeze-out time is suppressed,
  compared to the one at present.

As is well-known, the SM Higgs potential becomes unstable at high energies,
  since the running SM Higgs quartic coupling ($\lambda$ in Eq.~(\ref{H-Delta-Potential}))
  turns negative at the renormalization scale of $\mu = {\cal O}(10^{10})$ GeV \cite{Buttazzo:2013uya}.
However, it has been shown in Ref.~\cite{Gogoladze:2008gf} (before the Higgs boson discovery)
  that this electroweak vacuum instability can be solved in the presence of type II seesaw.
See Ref.~\cite{Dev:2013ff} for follow-up analysis after the Higgs boson discovery at the Large Hadron Collider.

In summary, motivated by the DAMPE cosmic ray $e^+ + e^-$ excess,
  we have revisited a simple extension of the SM
  to supplement it with neutrino masses via type II seesaw and
  a stable SM singlet scalar as the DM candidate.
With a suitable choice of the couplings among the DM particles,
  the Higgs doublet and the triplet of type II seesaw,
  the DM particles in our galactic halo annihilate into a pair of triplet scalars,
  and their subsequent decays produce high energy CREs.
Through the type II seesaw mechanism, the flavor structure of the primary leptons
  created by the triplet decay has a direct relation with the neutrino oscillation data.
We have found that the DM interpretation of the DAMPE excess determines
  the pattern of neutrino mass spectrum to be the inverted hierarchy type,
  taking into account the constraints from the Fermi-LAT observations of dwarf spheroidal galaxies.

%%%%%%%%%%%%%%%%%%%%%%%%%%%%%%%%%%
\section*{Acknowledgments}
%%%%%%%%%%%%%%%%%%%%%%%%%%%%%%%%%%
This work is supported in part by the DOE Grant No.~DE-SC0013680 (N.O.) and No.~DE-SC0013880 (Q.S.).

%\newpage
%%%%%%%%%%%%%%%%%%%%%%%%%%%%%%%%%

\end{document}